\documentclass[%
 aip,
 jmp,%
 amsmath,amssymb,
 reprint,%
balancelastpage,
]{revtex4-1}

\usepackage{graphicx}
\graphicspath{{./}{./images/}{../../images/}}

\usepackage{dcolumn}
\usepackage{bm}

\begin{document}

\preprint{AIP/123-QED}

\title{Comment on ``An error-resilient non-volatile magneto-elastic universal logic gate with ultralow energy-delay product'' [Sci. Rep. 4, 7553 (2014)]}

\author{Kuntal Roy}
\email{royk@purdue.edu.}
\noaffiliation
\affiliation{Department of Electrical and Computer Engineering, Purdue University, West Lafayette, Indiana 47907, USA}

\maketitle

An article~\cite{biswa14_3} by Biswas and the coauthors Atulasimha and Bandyopadhyay (referred as BA onwards) claimed to devise an error-resilient and ultralow energy universal logic gate NAND using piezoelectric-magnetostrictive heterostructures. Here it is pointed out that the basic idea behind such gate using piezoelectric-magnetostrictive heterostructures with multiple inputs is followed from Ref.~\onlinecite{roy13}. The Ref.~\onlinecite{biswa14_3} pertains to neither error-resilient nor energy-efficient operation, which follows from BA's own contention. Moreover, the statements made in the Ref.~\onlinecite{biswa14_3} on Ref.~\onlinecite{roy13} are incorrect and misleading. Furthermore, the proposal in Ref.~\onlinecite{biswa14_3} is highly area-consuming and lacks scalability, contrary to what is claimed in the article. There are other technical issues too in Ref.~\onlinecite{biswa14_3}, as pointed out here. 

First, note that Ref.~\onlinecite{biswa14_3} uses a $\sim90^\circ$ (precisely $86.4^\circ$) switching mechanism following an earlier idea by others~\cite{RefWorks:850,RefWorks:559,RefWorks:848,RefWorks:849}. Therefore, \emph{magnetization does not switch a complete $180^\circ$} in Ref.~\onlinecite{biswa14_3} and this leads to a low	tunneling magnetoresistance (TMR) while reading the magnetization state and it has serious consequence on \emph{read} error probability. The researchers are in fact trying to increase the TMR using half-metals (rather than using CoFeB) to avoid high read-error rate, so lowering the angular separation to $90^\circ$ (from $180^\circ$) is an important issue in Ref.~\onlinecite{biswa14_3}.

Next, BA's comments on  Ref.~\onlinecite{roy13}, which follows a complete $180^\circ$ switching mechanism~\cite{roy13_2,roy11_6}, are factually and technically incorrect and misleading, as explained here. BA state that it requires a sensing circuitry for operation of the proposal in the Ref.~\onlinecite{roy13} and indicates as if it is an issue with the computing proposal in Ref.~\onlinecite{roy13}. First, BA do not point out at all that the sensing element is required for complete $180^\circ$ switching only and it does not require so if $90^\circ$ switching mechanism is used, as BA have used in Ref.~\onlinecite{biswa14_3}. Second, Ref.~\onlinecite{roy13} just uses the existing complete $180^\circ$ magnetization switching methodology~\cite{roy11,roy11_2,roy13_2,roy11_6}.  In Ref.~\onlinecite{roy13}, it is clearly mentioned while referring the Ref.~\onlinecite{roy13_2} that ``Computing methodologies utilizing such 180$^\circ$ switching mechanism between the two stable states of a shape-anisotropic magnetostrictive nanomagnet have not been proposed so far.'' The computing proposal in Ref.~\onlinecite{roy13} is based on the switching methodology explained in the Ref.~\onlinecite{roy13_2} (and also Refs.~\onlinecite{roy11,roy11_2,roy11_6}) and it was stated in the Ref.~\onlinecite{roy13} clearly.

In Ref.~\onlinecite{roy13_comment}, BA comment that the sensing circuitry for magnetization switching (complete $180^\circ$) to dissuade thermal fluctuations ``will \emph{likely} require multiple charge-based electronic devices (e.g. transistors) that are known energy hogs'' is {diffident}, misses the {basic} understanding level, and {completely} contradicts the big picture involved. Note that researchers are trying to replace the traditional \emph{switch} based on charge-based transistors by a new possible ``ultra-low-energy'' \emph{switch} (e.g., using multiferroic composites). Therefore, any circuitry can be built with the energy-efficient switch itself rather than the conventional transistors. Usually, it requires several peripheral circuitry in conjunction with the basic switch in a system~\cite{rabae03,pedra02}.  While researchers report on the performance metrics of the basic switch itself, the total energy dissipation considering the other required circuitry does not change the order of energy dissipation, utilizing the respective devices~\cite{rabae03,pedra02}. This was the understanding while claiming energy-efficiency in Ref.~\onlinecite{roy11_6}, where BA are coauthors. Therefore, BA's contention that the computing methodology in Ref.~\onlinecite{roy13} cannot pertain to ``ultra-low-energy'' operation due to the underlying magnetization switching methodology is {untenable} and violates the ethical policy of coauthorship~\cite{AIP_ethics} (``Any individual unwilling or unable to accept appropriate responsibility for a paper should not be a coauthor.''). Moreover, BA conceived such understandings from Roy~\cite{roy13_2}, which makes the BA's comments~\cite{roy13_comment} devoid of any scientific reasoning in this respect. Several new ideas on the switching methodology exerting more asymmetry in the system may come along too, e.g., interface and exchange coupled systems, as evaluated in Ref.~\onlinecite{roy14_2} by Roy, which would not require any sensing circuitry and also it can maintain the direction of switching unlike toggle switching~\cite{roy13_2,biswa14}.

As stated in the Ref.~\onlinecite{roy13_2}, the sensing circuitry to dissuade thermal fluctuations can be implemented by measuring the magnetoresistance in a spin-valve or magnetic tunnel junction (MTJ), which needs to be used anyway to \emph{read} the magnetization state and it dissipates much less energy than for \emph{write} operation. We know the magnetoresistance of the MTJ when magnetization resides at nanomagnet's hard-plane and comparing this known signal with the sensed signal of the MTJ, the stress can be ramped down~\cite{roy13_2}. Such comparator can be implemented with these energy-efficient multiferroic devices, i.e., charge-based transistors do not need to be utilized. Moreover, the fabrication procedure of transistors and nanomagnets are different, therefore, it is beneficial to use the same device throughout rather than having layers of different devices with different fabrication procedures.

Note that there are inconsistencies in the Ref.~\onlinecite{biswa14_3} co-authored by BA while comparing their Comment~\cite{roy13_comment}. In their Comment~\cite{roy13_comment}, BA state ``charge-based electronic devices (e.g., transistors) that are known energy hogs'', however, note that in Ref.~\onlinecite{biswa14_3}, BA state that ``a low-power transistor may dissipate only $10^3$ kT of energy when it switches in 0.1 ns (energy-delay product = $3 \times 10^{-28}$ J-s)''. The key point is that there exists switching delay-energy trade-off in devices, i.e., if a device is switched slower, it dissipates a lower energy. Therefore, if transistors were to switch slower, the energy dissipation could have been much lower. Ref.~\onlinecite{roy13}, on which BA's Comment~\cite{roy13_comment} is based, pertains to switching delay of $\sim$1 ns, therefore even if transistors are utilized to build any additional hardware (however, this is a misconception as described above), then energy dissipation would not have been issue, contrary to BA's own claim.~\cite{roy13_comment} The challenge is to reduce energy dissipation of a device while keeping the switching delay intact~\cite{roy13_spin}. There can be other metrics like area and error-probability in the trade-off analysis too. BA's Comment~\cite{roy13_comment} did not take such trade-off  into consideration while defining transistors as energy hogs and commenting on Ref.~\onlinecite{roy13}. Anyway, as mentioned earlier, the charge-based transistors need not be used to build any additional hardware, rather multiferroic devices themselves can be utilized. 

Ironically, Ref.~\onlinecite{biswa14_2}, in which BA are coauthors, proposed a ``toggle'' switch (as stated that ``a write cycle must be preceded by a read cycle to determine the stored bit''), which would require a similar use of spin-valve or MTJ for reading the known bit, storing it, and then using it for \emph{comparison}. According to BA's own contention~\cite{roy13_comment}, such \emph{additional} circuitry needs to be constructed with energy-inefficient transistors, invalidating the claim of energy efficiency in Ref.~\onlinecite{biswa14_2}.  Also, Ref.~\onlinecite{munir14}, in which BA are coauthors, uses some pulse shaping methodology (which, however, leads to ``high'' error probability at high switching speed that is required to build general-purpose nanoelectronics~\cite{itrs,nri}). According to BA's own contention~\cite{roy13_comment}, the circuitry for generating the precisely shaped pulse needs to be constructed with energy-inefficient transistors, invalidating the claim of energy efficiency in Ref.~\onlinecite{biswa14_2}. (Note that one additional hardware cannot be shared between many devices distributed on a chip due to interconnect delay and loading effect.) Therefore, BA are contradicting their own Comment~\cite{roy13_comment}, however, {incorrect} and {misleading}.

Similarly, in Ref.~\onlinecite{biswa14_3}, according to BA's own contention~\cite{roy13_comment}, the current source $I_{BIAS}$ and resistors would need to be built with transistors invalidating the claim of energy efficiency in Ref.~\onlinecite{biswa14_3}. According to BA's contention~\cite{roy13_comment}, in a system, any circuitry other than the basic switch itself needs to be constructed with energy-inefficient transistors and in this way no system based on nanomagnetic logic would be energy-efficient. (Note that one current source $I_{BIAS}$ cannot be shared between many devices distributed on a chip due to interconnect delay and loading effect.) However, this is incorrect since BA missed the {basic} understanding level here as explained beforehand. Note that BA did not complain on such energy-inefficiency in papers on magnetization switching methodology~\cite{roy11_6} where they are coauthors.

BA talks about nanosecond switching speed of the device in Ref.~\onlinecite{biswa14_3}. Interestingly, such nanosecond switching delay cannot be conceived by considering only the in-plane potential landscapes of magnetization (Figs.~2 and~3 in Ref.~\onlinecite{biswa14_3}). Magnetization may deflect out of magnet's plane when stress is applied since the torque due to stress acts in the out-of-plane direction as 
\begin{equation}
\mathbf{T_{E,stress}} = - \mathbf{\hat{e}_r} \times \nabla\,E_{stress} = - (3/2) \, \lambda_s \sigma \Omega sin(2\theta)\, \mathbf{\hat{e}_\phi},
\label{eq:T_stress}
\end{equation}
where $E_{stress}=- (3/2) \, \lambda_s \sigma \Omega cos^2\theta$ is the potential energy due to stress, $(3/2)\lambda_s$ is the magnetostrictive coefficient, $\sigma$ is the stress, $\Omega$ is nanomagnet's volume, $\theta$ and $\phi$ are polar and azimuthal angles in spherical coordinate system [represented by ($r$,$\theta$,$\phi$)], respectively~\cite{roy13_spin}. Therefore, \emph{magnetization rotates out-of-plane} even if the demagnetization factor in the out-of-plane direction is high ($\sim$10 times compared to the in-plane directions). A slight out-of-plane excursion has important ramification to increase switching speed tremendously~\cite{roy13_2,roy11,roy11_2,roy11_6}, which cannot be conceived if magnetization is \emph{assumed} to reside always on magnet's plane as \emph{assumed} by BA in Ref.~\onlinecite{RefWorks:154} {unreasonably}. If we calculate the switching delay of magnetization according to Ref.~\onlinecite{RefWorks:154}, it will {incorrectly} come out as $\sim$1000 ns, which is clearly exorbitantly high for general-purpose applications~\cite{itrs,nri}. Therefore, Ref.~\onlinecite{RefWorks:154} would have been \emph{untenable to build general-purpose nanoelectronics}. If charge based transistors were to operate \emph{slow}, the energy dissipation would not have been an issue~\cite{rabae03,itrs,nri}. However, the analysis in Ref.~\onlinecite{RefWorks:154} is incorrect. Stress rotates magnetization out of magnet's plane and this generates a helpful torque that rotates magnetization \emph{fast}, increasing the switching speed to more than 1 GHz~\cite{roy11,roy11_2,roy13_2,roy11_6}. This was first \emph{corrected} by Roy in Refs.~\onlinecite{roy11} and~\onlinecite{fasha11}, which put the nanomagnetic memory and logic based on multiferroic composites on solid footing~\cite{roy13_spin}. Such key idea by Roy and associated papers~\cite{roy11,roy11_2,roy13_2,roy11_6,fasha11} have been utilized by BA in a patent~\cite{roy12_patent}. But, BA are complaining~\cite{roy13_comment,biswa14_3} over the very same magnetization switching methodology used in Ref.~\onlinecite{roy13} by Roy.

There are two more misleading statements in Ref.~\onlinecite{biswa14_3} on Ref.~\onlinecite{roy13}. In Ref.~\onlinecite{roy13_comment}, BA agreed that with the help of a sensing circuitry, the error-resiliency can be achieved for the proposal in Ref.~\onlinecite{roy13}, which was already lucidly explained by Roy in Ref.~\onlinecite{roy13_2}. However, in Ref.~\onlinecite{biswa14_3}, BA comment that the proposal in Ref.~\onlinecite{roy13} is error-prone \emph{as well as} mentioning of the sensing circuitry needed for error-resilient $180^\circ$ switching~\cite{roy13_2}. The sensing circuitry is used for error-resilient switching.~\cite{roy13_2} Therefore, stating both sensing circuitry and error-prone (in Ref.~\onlinecite{biswa14_3} by BA) is categorically incorrect and {misleading}. 

Also, BA comment in Ref.~\onlinecite{biswa14_3} that the design in Ref.~\onlinecite{roy13} is flawed while referring their Comment~\cite{roy13_comment}, which is factually and technically incorrect and misleading. On design issues, BA raised a couple of {incorrect} issues on Ref.~\onlinecite{roy13} in their Comment~\cite{roy13_comment}. First, BA argue that the stresses generated by the two inputs do not add in magnitude. This is \emph{incorrect} and \emph{misleading}. The inputs generate \emph{strain} in the piezoelectric layer and each input generates a same strain. (Note that the inputs are \emph{symmetrically} placed on the piezoelectric layer.) The addition of strains due to two inputs can be simply understood from the superposition principle for a linear system (strain is proportional to the electric field~\cite{roy11,roy11_2,roy11_6,roy13_2}). Any detailed solver with underlying detailed equations can confirm that too~\cite{comsol}. The response of the system is strain and electric field is the input to the system. One should not consider the charge in the Poisson's equation since these are \emph{strain-mediated} multiferroic composites~\cite{roy13,roy14,roy11,roy11_2,roy11_6,roy13_2} and not the charge-mediated ones.

Second, BA raised an issue on the concatenation between the individual devices and that is {incorrect} and {misleading} too. Note that concatenation in Ref.~\onlinecite{roy13} is addressed and it is clearly mentioned that ``The SET operation precedes the LOGIC operation ...'' which BA did not take into account. For an individual gate, it needs to perform a SET operation \emph{before} going for LOGIC operation \emph{on that gate}. A voltage on the ``Set'' terminal is applied to perform the SET operation, which is stated in the paper. BA are \emph{not} considering this (note that there are no SET inputs in the Fig.~1 of the Comment by BA~\cite{roy13_comment}) and hence this is a factually {incorrect} and {misleading} point raised by BA.

Note that Ref.~\onlinecite{biswa14_3} has a very high \emph{read} error probability, which is a consequence of using a $90^\circ$ switching mechanism, contrary to complete $180^\circ$ switching mechanism used in Ref.~\onlinecite{roy13}. Following Ref.~\onlinecite{biswa14_3}, in the high logic state, a 5\% variation of input voltage is intolerable and fails the logic operation. Therefore the error-resiliency claim in the Ref.~\onlinecite{biswa14_3} is {severely flawed}.

Also, the claim and notion of scalability discussed in Ref.~\onlinecite{biswa14_3} are flawed. Equation~\eqref{eq:T_stress} clearly says that the stress anisotropy is proportional to the volume of the nanomagnet. This is similar to magnetic field based switching, i.e., at a lower volume of the nanomagnet, it requires a higher stress (and a higher voltage that generates the stress) to produce the same stress anisotropy. Unless additional strategies are incorporated, the claim of scalability in Ref.~\onlinecite{biswa14_3} is {untenable}. The other properties stated for logic operation in Ref.~\onlinecite{biswa14_3} are \emph{not} new in literature. 

In Ref.~\onlinecite{fasha13} (which is referred in Ref.~\onlinecite{biswa14_3}), BA predicated the {demise of Bennett clocking mechanism} in the presence of room-temperature thermal fluctuations saying ``This could render nanomagnetic logic schemes that rely on dipole coupling to perform Boolean logic operations.'' But the critical analysis performed in Ref.~\onlinecite{roy14} says otherwise. Ref.~\onlinecite{roy14} showed that the out-of-plane excursion of magnetization (which was {missed} by BA in Ref.~\onlinecite{RefWorks:154}) combined with the thermal fluctuations is the reason behind switching failures and BA failed to grasp such understanding in Ref.~\onlinecite{fasha13} and thereby {incorrectly} predicting the demise of general-purpose nanomagnetic logic. Note that Ref.~\onlinecite{munir14}, in which BA are coauthors, is error-prone at high switching speed and therefore the operation at low switching speed ($\sim$10 MHz) is untenable for building general-purpose nanoelectronics, while for niche applications it still needs to compete with the existing transistor based technology, which is energy-efficient at low switching speed.

Note that Ref.~\onlinecite{roy13}, on which BA's Comment~\cite{roy13_comment} is based, presents a novel intriguing methodology of building logic rather than Bennett clocking mechanism using multiferroic composites~\cite{roy13_spin,fasha11}, i.e., using a single device with a read-unit (MTJ) as a \emph{switch} (similar to that a transistor acts as a \emph{switch}). Also, it is shown in Ref.~\onlinecite{roy13} how to increase the \emph{functionality} per device, e.g., it proposed universal logic gates (NAND and NOR) utilizing a \emph{single} device with the well-established concept of using multiple contacts on the device to add up the strains generated in piezoelectric~\cite{comsol}, and a \emph{Set} input to preset the non-volatile magnetization state and facilitate concatenation, which are not conceived by BA in their Comment~\cite{roy13_comment}. Note that it is advantageous to increase the \emph{functionality} per \emph{single} device since stress anisotropy generated in the magnetostrictive nanomagnets is proportional to nanomagnet's volume and therefore, these \emph{single} multiferroic devices are area-inefficient. Hence, the proposal in Ref.~\onlinecite{roy13} can facilitate a highly-dense yet an ultra-low-energy computing paradigm. In Ref.~\onlinecite{biswa14_3}, the authors follow the same principles as in Ref.~\onlinecite{roy13}.

Ref.~\onlinecite{biswa14_3} uses resistors and potential divider (see Fig.~1 in Ref.~\onlinecite{biswa14_3}) to accommodate multi-inputs. The resistors need to be implemented additionally and would consume area. Comparatively, the device design in Ref.~\onlinecite{roy13} uses \emph{intrinsic} strain-addition property of piezoelectrics. The \emph{external} manipulation of inputs in Ref.~\onlinecite{biswa14_3} is a matter of concern. Note that such external way of using multi-inputs can be also used for traditional transistors to build universal logic gates, but it is not done, rather multiple transistors are used. It is stated in Ref.~\onlinecite{biswa14_3} (in the supplementary material) that ``The dissipation in the resistance $R$ can be negligible as we can make this resistance arbitrarily high.'' This is incorrect (any standard electrical engineering undergraduate textbook can be consulted) since $RC$ delay may be too high. Therefore, such design proposed in Ref.~\onlinecite{biswa14_3} is untenable.

There are some other technical flaws in Ref.~\onlinecite{biswa14_3} too. Ref.~\onlinecite{biswa14_3} says that the hard nanomagnet in the fixed layer is \emph{not} magnetostrictive. However, the hard nanomagnets in the fixed layer (synthetic antiferromagnetic layer) are usually made of CoFeB, which is magnetostrictive. Also, stress is {unreasonably assumed} to be removed abruptly in Ref.~\onlinecite{biswa14_3}, however, it was shown by Roy~\cite{roy13_2,roy11_6} that finite ramp rate of stress has immense consequence on magnetization dynamics particularly it can cause switching failures in the presence of thermal fluctuations. Also, Ref.~\onlinecite{biswa14_3} {incorrectly} states that the resistance changes by \emph{twice} using a wrong equation~\cite{biswa14} (while additionally \emph{assuming} 100\% spin injection/detection efficiencies) for the $90^\circ$ switching mechanism used therein. Moreover, Ref.~\onlinecite{biswa14_3} {incorrectly} calculated the demagnetization factors and the energy barrier height for a nanomagnet with the assumption of major axis (a)/minor axis (b)$\sim 1$ (while a/b=100 nm/42 nm), which is {very unreasonable}. Furthermore, the design proposed in Ref.~\onlinecite{biswa14_3} (also in Refs.~\onlinecite{biswa14,biswa14_2}) is highly area-consuming due to using lateral piezoelectric pads. There is an ongoing drive to reduce the area-consumption~\cite{RefWorks:774,roy14_2}, but Refs.~\onlinecite{biswa14_3,biswa14,biswa14_2} took such drive {backwards}.

Note that Ref.~\onlinecite{biswa14} (the authors are same as of in Ref.~\onlinecite{biswa14_3}) claimed a superior design of magnetoelastic memory ({incomplete non-$180^\circ$ switching of magnetization}), compared to an earlier idea~\cite{RefWorks:850,RefWorks:559,RefWorks:848,RefWorks:849}. The results presented in Ref.~\onlinecite{biswa14} are, however, {incorrect} and actually the switching delay and error-probability are inferior to the earlier idea~\cite{RefWorks:850,RefWorks:559,RefWorks:848,RefWorks:849}. In the Ref.~\onlinecite{biswa14_3}, note that BA have utilized the magnetization switching methodology as in the Refs.~\onlinecite{RefWorks:850,RefWorks:559,RefWorks:848,RefWorks:849}. If Ref.~\onlinecite{biswa14} performs better than Refs.~\onlinecite{RefWorks:850,RefWorks:559,RefWorks:848,RefWorks:849}, then BA simply could have utilized the switching methodology in Ref.~\onlinecite{biswa14} for the logic design in Ref.~\onlinecite{biswa14_3}.

Finally, note that Fig.~4 in the Ref.~\onlinecite{biswa14_3} showing gain in the system is already there in literature. Ref.~\onlinecite{roy_aps_2014} first states about the voltage amplification, i.e., gain with the procedure to evaluate that. Such characteristics (Fig.~4 in the Ref.~\onlinecite{biswa14_3}) has been already reported in literature (see Fig.~4(b) in Ref.~\onlinecite{roy_mrs_2014} and Fig.~5(b) in Ref.~\onlinecite{roy_spie_2014}). However, Ref.~\onlinecite{biswa14_3} does not mention the references~\cite{roy_aps_2014,roy_mrs_2014,roy_spie_2014}.

To summarize, the central claims and analysis of the Ref.~\onlinecite{biswa14_3}, while following Ref.~\onlinecite{roy13}, are {flawed}. The \emph{incomplete} $90^\circ$ switching mechanism is an issue behind the high \emph{read} error rate in Ref.~\onlinecite{biswa14_3}. Ref.~\onlinecite{biswa14_3} provides \emph{misleading} statements on Ref.~\onlinecite{roy13}, which uses a complete $180^\circ$ switching methodology. BA have repeatedly failed to conceive the key understandings, e.g., magnetization is {unreasonably} \emph{assumed} by BA to be confined on magnet's plane~\cite{RefWorks:154}, which has key consequence on magnetization dynamics underestimating the switching speed tremendously~\cite{roy13_spin,roy13_2} and causing issues in regards to error probability at room-temperature analysis~\cite{roy14}. Incorrectly, BA have also predicted {the demise of nanomagnetic logic}~\cite{fasha13}. And, most recently, in Ref.~\onlinecite{biswa14_3}, BA have also come up with {misleading} statements on Ref.~\onlinecite{roy13}, as explained here. Both theoretical and experimental efforts are emerging in this field of research on further improving the performance metrics and the \emph{comprehensive} discussion clearing the facts here would hopefully play an important role in possibly devising the magnetization switching methodology and nanomagnetic logic.


\begin{thebibliography}{10}
\expandafter\ifx\csname urlstyle\endcsname\relax
  \providecommand{\doi}[1]{doi:\discretionary{}{}{}#1}\else
  \providecommand{\doi}{doi:\discretionary{}{}{}\begingroup
  \urlstyle{rm}\Url}\fi

\bibitem{biswa14_3}
Biswas, A.~K., Bandyopadhyay, S. \& Atulasimha, J.
\newblock An error-resilient non-volatile magneto-elastic universal logic gate
  with ultralow energy-delay product.
\newblock \emph{Sci. Rep.} \textbf{4}, 7553 (2014).

\bibitem{roy13}
Roy, K.
\newblock Ultra-low-energy non-volatile straintronic computing using single
  multiferroic composites.
\newblock \emph{Appl. Phys. Lett.} \textbf{103}, 173110 (2013).

\bibitem{RefWorks:850}
Tiercelin, N., Dusch, Y., Preobrazhensky, V. \& Pernod, P.
\newblock Magnetoelectric memory using orthogonal magnetization states and
  magnetoelastic switching.
\newblock \emph{J. Appl. Phys.} \textbf{109}, 07D726 (2011).

\bibitem{RefWorks:559}
Tiercelin, N. \emph{et~al.}
\newblock Room temperature magnetoelectric memory cell using stress-mediated
  magnetoelastic switching in nanostructured multilayers.
\newblock \emph{Appl. Phys. Lett.} \textbf{99}, 192507 (2011).

\bibitem{RefWorks:848}
Giordano, S., Dusch, Y., Tiercelin, N., Pernod, P. \& Preobrazhensky, V.
\newblock Combined nanomechanical and nanomagnetic analysis of magnetoelectric
  memories.
\newblock \emph{Phys. Rev. B} \textbf{85}, 155321 (2012).

\bibitem{RefWorks:849}
Giordano, S., Dusch, Y., Tiercelin, N., Pernod, P. \& Preobrazhensky, V.
\newblock Thermal effects in magnetoelectric memories with stress-mediated
  switching.
\newblock \emph{J. Phys. D: Appl. Phys.} \textbf{46}, 325002 (2013).

\bibitem{roy13_2}
Roy, K., Bandyopadhyay, S. \& Atulasimha, J.
\newblock Binary switching in a `symmetric' potential landscape.
\newblock \emph{Sci. Rep.} \textbf{3}, 3038 (2013).

\bibitem{roy11_6}
Roy, K., Bandyopadhyay, S. \& Atulasimha, J.
\newblock Energy dissipation and switching delay in stress-induced switching of
  multiferroic nanomagnets in the presence of thermal fluctuations.
\newblock \emph{J. Appl. Phys.} \textbf{112}, 023914 (2012).

\bibitem{roy11}
Roy, K., Bandyopadhyay, S. \& Atulasimha, J.
\newblock Hybrid spintronics and straintronics: A magnetic technology for ultra
  low energy computing and signal processing.
\newblock \emph{Appl. Phys. Lett.} \textbf{99}, 063108 (2011).

\bibitem{roy11_2}
Roy, K., Bandyopadhyay, S. \& Atulasimha, J.
\newblock Switching dynamics of a magnetostrictive single-domain nanomagnet
  subjected to stress.
\newblock \emph{Phys. Rev. B} \textbf{83}, 224412 (2011).

\bibitem{roy13_comment}
Bandyopadhyay, S. \& Atulasimha, J.
\newblock Comment on ``{Ultra}-low-energy non-volatile straintronic computing
  using single multiferroic composites'' [{Appl. Phys. Lett.} 103, 173110
  (2013)].
\newblock \emph{Appl. Phys. Lett.} \textbf{105}, 176101 (2014).
\newblock Under investigation.

\bibitem{rabae03}
Rabaey, J.~M., Chandrakasan, A.~P. \& Nikoli\c{c}, B.
\newblock Digital Integrated Circuits.
\newblock Pearson Education (2003).

\bibitem{pedra02}
Pedram, M. \& Rabaey, J.~M., eds.
\newblock Power aware design methodologies.
\newblock Kluwer Academic Publishers (2002).

\bibitem{AIP_ethics}
\emph{See http://publishing.aip.org/publishing/authors/ethics for AIP's ethical
  policy of coauthorship} .

\bibitem{roy14_2}
Roy, K.
\newblock Electric field-induced magnetization switching in interface-coupled
  multiferroic heterostructures: a highly-dense, non-volatile, and
  ultra-low-energy computing paradigm.
\newblock \emph{J. Phys. D: Appl. Phys.} \textbf{47}, 252002 (2014).

\bibitem{biswa14}
Biswas, A.~K., Bandyopadhyay, S. \& Atulasimha, J.
\newblock Energy-efficient magnetoelastic non-volatile memory.
\newblock \emph{Appl. Phys. Lett.} \textbf{104}, 232403 (2014).

\bibitem{roy13_spin}
Roy, K.
\newblock Ultra-low-energy straintronics using multiferroic composites.
\newblock \emph{SPIN} \textbf{3}, 1330003 (2013).

\bibitem{biswa14_2}
Biswas, A.~K., Bandyopadhyay, S. \& Atulasimha, J.
\newblock Complete magnetization reversal in a magnetostrictive nanomagnet with
  voltage-generated stress: A reliable energy-efficient non-volatile
  magneto-elastic memory.
\newblock \emph{Appl. Phys. Lett.} \textbf{105}, 072408 (2014).

\bibitem{munir14}
Munira, K. \emph{et~al.}
\newblock Reducing error rates in straintronic multiferroic dipole-coupled
  nanomagnetic logic by pulse shapingy.
\newblock \emph{arXiv1405.4000v1}  (2014).

\bibitem{itrs}
http://www.itrs.net.

\bibitem{nri}
http://www.src.org/program/nri/.

\bibitem{RefWorks:154}
Atulasimha, J. \& Bandyopadhyay, S.
\newblock Bennett clocking of nanomagnetic logic using multiferroic
  single-domain nanomagnets.
\newblock \emph{Appl. Phys. Lett.} \textbf{97}, 173105--1--173105--3 (2010).

\bibitem{fasha11}
Fashami, M.~S., Roy, K., Atulasimha, J. \& Bandyopadhyay, S.
\newblock Magnetization dynamics, Bennett clocking and associated energy
  dissipation in multiferroic logic.
\newblock \emph{Nanotechnology} \textbf{22}, 155201 (2011).

\bibitem{roy12_patent}
Atulasimha, J. \& Bandyopadhyay, S.
\newblock Planar Multiferroic/Magnetostrictive Nanostructures as Memory
  Elements, Two-Stage Logic Gates and Four-State Logic Elements for Information
  Processing (2012).
\newblock {US} Patent 20120267735.

\bibitem{comsol}
http://www.comsol.com.

\bibitem{roy14}
Roy, K.
\newblock Critical analysis and remedy of switching failures in straintronic
  logic using Bennett clocking in the presence of thermal fluctuations.
\newblock \emph{Appl. Phys. Lett.} \textbf{104}, 013103 (2014).

\bibitem{fasha13}
Fashami, M.~S., Munira, K., Bandyopadhyay, S., Ghosh, A.~W. \& Atulasimha, J.
\newblock Switching of Dipole Coupled Multiferroic Nanomagnets in the Presence
  of Thermal Noise: Reliability of Nanomagnetic Logic.
\newblock \emph{IEEE Trans. Nanotechnol.} \textbf{12}, 1206--1212 (2013).

\bibitem{RefWorks:774}
Ikeda, S. \emph{et~al.}
\newblock A perpendicular-anisotropy CoFeB--MgO magnetic tunnel junction.
\newblock \emph{Nature Mater.} \textbf{9}, 721--724 (2010).

\bibitem{roy_aps_2014}
Roy, K.
\newblock Ultra-low-energy analog straintronics using multiferroic composites.
\newblock In American Physical Society (APS) March 2014 Meeting, Denver,
  Colorado, Mar 3, Session A8.6 (2014).

\bibitem{roy_mrs_2014}
Roy, K.
\newblock Ultra-low-energy straintronics using multiferroic composites.
\newblock In Materials Research Society (MRS) Spring 2014 Meeting, Mater. Res.
  Soc. Symp. Proc. 1691 (2014).

\bibitem{roy_spie_2014}
Roy, K.
\newblock Ultra-low-energy straintronics using multiferroic composites.
\newblock In Proc. SPIE Nanoscience (Spintronics VII) 9167, 91670U (2014).

\end{thebibliography}

\end{document}


\title[]{Comment on ``Reducing error rates in straintronic multiferroic nanomagnetic logic by pulse shaping'' [Nanotechnology 26, 245202 (2015)]}

\author{Kuntal Roy}
\address{School of Electrical and Computer Engineering, Purdue University, West Lafayette, Indiana 47907, USA}

\ead{royk@purdue.edu}

\bigskip

In Ref.~\citenum{munir15}, the authors proposed to reduce error probability of switching in a system of \emph{two} dipole-coupled magnetostrictive nanomagnets in strain-mediated multiferroic heterostructures using voltage (stress) pulse shaping. The authors conclude that high success probability of switching \emph{cannot} be achieved at high switching speed ($\sim$1 ns), and therefore their proposed system is only applicable for niche applications. However, such conclusion lacked the critical understandings behind such high error probability for general-purpose logic applications. Fortunately, such analysis and a possible solution along the line of the analysis is present in literature~\cite{roy14} using Bennett clocking~\cite{RefWorks:144} for logic design, on which Ref.~\citenum{munir15} (and arXiv version Ref.~\citenum{munir14}) has made some \emph{incorrect} statements. Earlier, a subset of the authors of Ref.~\citenum{munir15} also published a paper~\cite{fasha13} on a \emph{four} magnet system using the Bennett clocking mechanism, where the authors also came up with a similar conclusion of high error probability and the demise of multiferroic nanomagnetic logic, however, without relevant analysis similar to the case as in the Ref.~\citenum{munir15}.

\begin{figure*}[t]
\centering
\includegraphics[width=\textwidth]{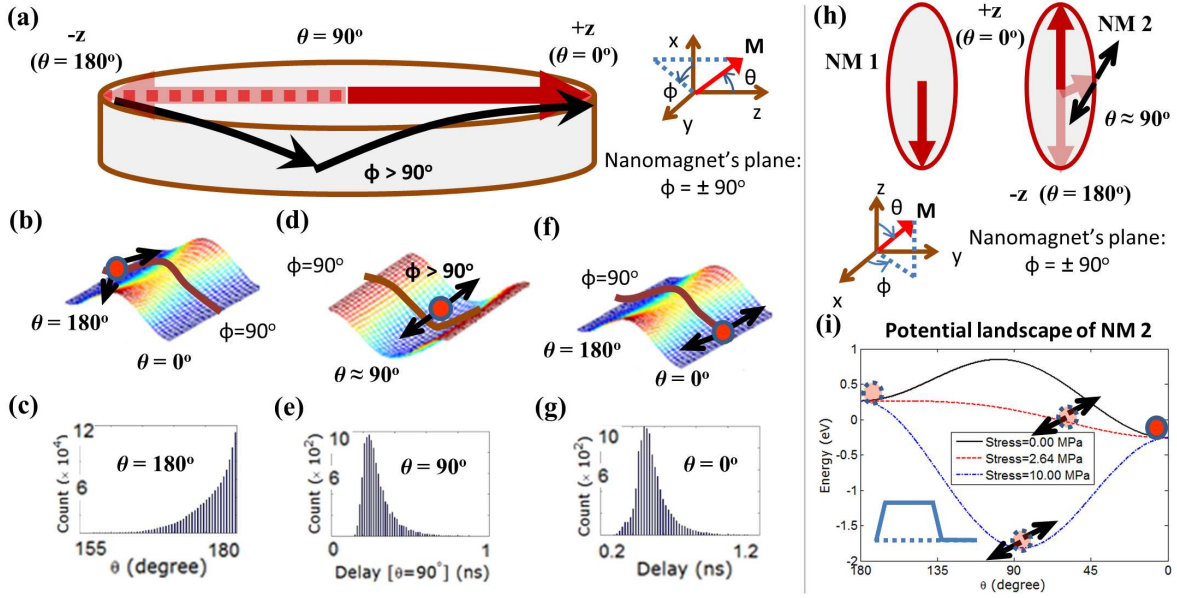}
\caption{\label{fig:Nanotechnology_26_245202_2015}
(a) The out-of-plane excursion of magnetization and axis assignment. 
(b), (c) Three-dimensional potential landscape of the nanomagnet and the distribution of magnetization in the presence of room-temperature (300 K) thermal fluctuations. The solid line depicts the \emph{in-plane} potential landscape.
(d), (e) The same as (b), (c) when magnetization reaches $\theta=90^\circ$.
(f), (g) The same as (b), (c) when magnetization reaches $\theta\simeq180^\circ$.
[parts (c), (e), and (g) are taken from Refs.~\citenum{roy11_5,roy13_2}.]
(h) Two-magnet system depicting that the second nanomagnet (NM 2) deflects out-of-plane when reaching $\theta \simeq 90^\circ$ upon application of stress.
(i) Potential landscapes of NM 2 with stress as a parameter depicting the successful traversal of magnetization from $\theta \simeq 180^\circ$ to $\theta \simeq 0^\circ$ upon application and withdrawal of stress.
}
\end{figure*}

First, the Fig.~1 in Ref.~\citenum{munir15} shows the directions of unit vectors in standard Cartesian coordinate system ($x$-$y$-$z$) incorrectly (one correct option is to interchange $x$ and $y$), and also the demagnetization factors in the Equations~(7),~(8), and~(9) are incorrect. This may have produced incorrect results in the paper~\cite{munir15}. In any case, a correct coordinate system will be utilized in this Comment as shown in Fig.~\ref{fig:Nanotechnology_26_245202_2015}(a). 

The Fig.~2(c) in Ref.~\citenum{munir15} concerns about the \emph{in-plane} potential landscape of the nanomagnets to analyze the error probability, however, the critical analysis in Ref.~\citenum{roy14} has shown that it is the \emph{out-of-plane} motion that is responsible for high error probability and a solution along the line of such critical analysis was presented. The intriguing dynamics of a complete $180^\circ$ switching for a single magnetostrictive nanomagnet has been studied elaborately in Refs.~\citenum{roy11_5,roy13_2}. The torque due to stress acts in the out-of-plane ($\mathbf{\hat{e}_\phi}$ in Fig.~\ref{fig:Nanotechnology_26_245202_2015}) direction as 
\begin{equation}
\mathbf{T_{E,stress}} = - \mathbf{\hat{e}_r} \times \nabla\,E_{stress} = - (3/2) \, \lambda_s \sigma sin(2\theta)\, \mathbf{\hat{e}_\phi}, \nonumber
\label{eq:T_stress}
\end{equation}
where $E_{stress}=- (3/2) \, \lambda_s \sigma cos^2\theta$ is the potential energy due to stress per unit volume, $(3/2)\lambda_s$ is the magnetostrictive coefficient, and $\sigma$ is the stress. If such out-of-plane motion is \emph{not} considered, the complete $180^\circ$ motion cannot be possible~\cite{roy11_5,roy13_2}. Also such switching is confirmed by others~\cite{RefWorks:872,RefWorks:873,RefWorks:874}. Such out-of-plane motion also increases the switching speed to the order of GHz~\cite{roy11_5,roy13_2}. 

Ref.~\citenum{RefWorks:154} \emph{incorrectly} assumed only \emph{in-plane} motion of magnetization for logic design using Bennett clocking, and the switching delay of magnetization according to Ref.~\citenum{RefWorks:154} is $\sim$1000 ns, and that is clearly \emph{untenable} for building general purpose nanoelectronics~\cite{itrs,nri}. Fortunately, the analysis in Ref.~\citenum{RefWorks:154} is incorrect and it was corrected by Roy for memory design~\cite{roy13_spin,roy11,roy11_2,roy11_5,roy11_6,roy13_2} and also for logic design~\cite{roy13_spin,fasha11,nano_edi}. Similarly,  in the presence of thermal fluctuations, the critical analysis due to \emph{out-of-plane} excursion of magnetization is missing in Refs.~\citenum{munir15,munir14,fasha13} for logic design purposes with wrong conclusions of high switching delay and high error probability; correct analysis has been performed by Roy in Ref.~\citenum{roy14}. Such out-of-plane motion also plays important role in determining the pertinent energy dissipation~\cite{roy10,biswa13,arXiv:1501.05941v1}.

Ref.~\citenum{munir15} has made the following \emph{incorrect} statements on Ref.~\citenum{roy14}, as explained below.

\begin{enumerate}

\renewcommand{\theenumi}{\arabic{enumi}}

	\item Ref.~\citenum{munir15} states while referring Ref.~\citenum{roy14} that ``it examined switching errors in straintronic memory (not logic)''. This is preposterous since Fig.~1(b) in Ref.~\citenum{roy14} clearly shows the logic design using Bennett clocking for four-nanomagnet system and the underlying equations are also derived and provided. Actually, Ref.~\citenum{munir15} analyzed only \emph{two}  nanomagnet system, therefore, the system in Ref.~\citenum{munir15} is \emph{incomplete} to be termed as logic design. 
	
	\item  Ref.~\citenum{munir15} states while referring Ref.~\citenum{roy14} that ``A feedback circuit will determine when the magnetization has rotated by $90^\circ$ ... at that juncture.'' This has been already stated and explained in literature~\cite{roy11_5,roy13_2} and Ref.~\citenum{munir15} stated it \emph{quite inaccurately}. Fig.~\ref{fig:Nanotechnology_26_245202_2015}(a) shows that magnetization gets deflected out-of-plane due to the torque exerted on it [see Equation~\eref{eq:T_stress}]. The three-dimensional potential landscapes are also shown in the Figs.~\ref{fig:Nanotechnology_26_245202_2015}(b), (d), (f), depicting the out-of-plane motion of magnetization. Due to thermal fluctuations, the initial angle distribution of magnetization [Fig.~\ref{fig:Nanotechnology_26_245202_2015}(c)] and the time it takes to reach $\theta=90^\circ$ is a \emph{wide} distribution [Fig.~\ref{fig:Nanotechnology_26_245202_2015}(e)]. Stress needs to be withdrawn when magnetization reaches \emph{around} $\theta=90^\circ$ so that magnetization traverses towards $\theta\simeq 0^\circ$ [switching delay distribution is shown in Fig.~\ref{fig:Nanotechnology_26_245202_2015}(g)], and does not backtrack towards $\theta\simeq 180^\circ$, causing switching failure~\cite{roy11_5,roy13_2}. This is purely dynamical phenomenon~\cite{roy11_5,roy13_2} contrary to stead-state analysis. Therefore, a sensing element is required to detect when magnetization reaches around $\theta=90^\circ$~\cite{roy11_5,roy13_2}. The sensing element can be a spin valve or magnetic tunnel junction (MTJ), which is required to read the magnetization state~\cite{roy11_5,roy13_2,roy15_1}. We can get calibrated on the magnetoresistance of the spin valve or MTJ when $\theta=90^\circ$ and comparing this known signal with the sensed signal of the MTJ, the stress can be ramped down~\cite{roy11_5,roy13_2}. According to the study presented in Refs.~\citenum{roy11_5,roy13_2}, note that there is tolerance around $\theta=90^\circ$, i.e., stress does not need to be withdrawn precisely \emph{at the juncture} $\theta=90^\circ$ (this was also \emph{incorrectly} stated in Refs.~\citenum{biswa13,biswa14,biswa14_2}).
	
	\hspace*{5mm} Similarly, in a \emph{two} nanomagnet system~\cite{munir15}, or in a complete \emph{four} nanomagnet system using Bennett clocking~\cite{roy14}, the out-of-plane excursion of magnetization needs to be considered as depicted in the Fig.~\ref{fig:Nanotechnology_26_245202_2015}(h). If magnetization gets deflected in so-called \emph{bad} $\phi$-quadrants, as explained in Refs.~\citenum{roy11_5,roy13_2,roy14}, the dipole coupling from the nanomagnet~1 [Fig.~\ref{fig:Nanotechnology_26_245202_2015}(h)] is not sufficient to switch the nanomagnet~2 in the correct direction, causing error in logic design. Ref.~\citenum{munir15} considers only \emph{in-plane} potential landscape while explaining, therefore it could not explain the primary reason behind the high error probability in magnetization switching for logic design purposes.
	
	\hspace*{5mm} It should be noted that it is possible to harness more asymmetry in the system, rather than depending only on the out-of-plane excursion of magnetization~\cite{roy11_5,roy13_2,roy14}, so that the sensing procedure and dynamic withdrawal of stress will not be required. As shown in Ref.~\citenum{roy14_2}, the interface and exchange coupling can create asymmetry in the system facilitating non-toggle switching, without any requirement of the sensing procedure.
	
	\item  Ref.~\citenum{munir15} uses a vague term `interactive pulse timing' regarding Ref.~\citenum{roy14}. As explained in the above point, the strategy proposed in Ref.~\citenum{roy14} is in line to the physical operation of the basic device operation, which Ref.~\citenum{munir15} could not conceive. The pulse shaping methodology proposed in Ref.~\citenum{munir15} is based on the understanding of \emph{in-plane} motion and \emph{not} on out-of-plane motion. Only in the very end of the paper, it vaguely states that ``provided the out-of-plane magnetization effects do not begin to dominate first.'' As explained in the above point, the contents in Ref.~\citenum{munir15} lack the key understandings and analysis behind the critical device operation and high error probability.
	
 \item Ref.~\citenum{munir15} states while referring Ref.~\citenum{roy14} that ``Unfortunately, the feedback circuit will dissipate so much energy that it defeats the very purpose of SML. It is therefore an ineffective countermeasure.'' Actually, the research presented in Ref.~\citenum{munir15} is an ineffective countermeasure since it concludes that low error probability cannot be achieved for logic design with their proposal, i.e., pulse shaping. Moreover, it could not analyze the reasoning behind high error probability, which nonetheless exists in the literature~\cite{roy14}. 

\hspace*{5mm} Ref.~\citenum{munir15} foretell \emph{without any reasoning} that the sensing element will dissipate too much energy. It appears that the authors in Ref.~\citenum{munir15} think that logic is limited to four magnet systems~\cite{fasha13} and it would require charge-based transistors to build any other circuitry~\cite{roy13,roy13_comment,arxiv:1412.2200v1,arXiv:1412.8244v1,biswa14_3}. Such thinking misses the big picture involved. Note that researchers are trying to replace the traditional transistors by a new possible energy-efficient switch (e.g., using multiferroic composites). Therefore, any required circuitry in general can be built with the energy-efficient switch itself rather than using charge-based transistors. Usually, it requires several peripheral circuitry in addition to the basic switch itself and it does not change the orders of energy dissipation~\cite{rabae03,pedra02}.

\hspace*{5mm} As stated in Refs.~\citenum{roy11_5,roy13_2}, the sensing element to detect when magnetization reaches \emph{around} $\theta=90^\circ$ can be implemented with spin valve or magnetic tunnel junction, which is anyway required to read the magnetization state of a nanomagnet~\cite{roy15_1}. And it is well-known that a \emph{small} magnitude of current is required to read the magnetization state, leading to only \emph{miniscule energy dissipation}. Therefore, the comment made by Ref.~\citenum{munir15} is untenable. 

\hspace*{5mm} A few contradicting facts vis-a-vis the comment made by Ref.~\citenum{munir15} are pointed out below.
	
	\begin{enumerate}

		\item Ironically, a couple of authors in Ref.~\citenum{munir15}, Bandyopadhyay and Atulasimha (referred as BA onwards) are coauthors of Roy in Refs.~\citenum{roy11,roy11_2,roy11_5,roy11_6,roy13_2}. In particular, energy efficiency is claimed in the presence of thermal fluctuations in Ref.~\citenum{roy11_6}, which requires the sensing element too. Also, there is a patent~\cite{roy12_patent} filed by BA including Roy's contributions~\cite{roy11,roy11_2,roy11_5,roy11_6,roy13_2,roy13_spin,fasha11} claims energy-efficiency requiring the sensing element therein as well. Therefore, the comment made by Ref.~\citenum{munir15} is very perplexing.

		\item Note that Ref.~\citenum{munir15} uses precisely shaped pulses. Such pulses need to be generated too using some circuitry. According to the perception of Ref.~\citenum{munir15}, transistors need to be utilized and the system would dissipate too much energy, invalidating the claim of energy efficiency in Ref.~\citenum{munir15}. Note that one additional hardware cannot be shared between many devices distributed on a chip due to interconnect delay and loading effect. Also, note that pulse shaping is an ineffective countermeasure since it is not helping much in reducing the error probability, therefore building and using such circuitry do not make sense.
		
		\item Ref.~\citenum{biswa14_2}, in which BA are coauthors, proposed a ``toggle'' switch (as stated that ``a write cycle must be preceded by a read cycle to determine the stored bit''), which would require a similar use of spin-valve or MTJ for reading the known bit, storing it, and then using it for \emph{comparison}.  According to the perception of Ref.~\citenum{munir15}, such \emph{additional} circuitry needs to be constructed with energy-inefficient transistors, invalidating the claim of energy efficiency in Ref.~\citenum{biswa14_2}. 

	\end{enumerate}

	Fortunately, the perception in Ref.~\citenum{munir15} is \emph{incorrect}, otherwise, no system comprised of multiferroic devices would have been energy-efficient; transistors would have been required always to build the peripheral circuitry. Note that such sensing circuitry is not always required. It is shown in Ref.~\citenum{roy14_2} that the interface and exchange coupling can create enough asymmetry in the system facilitating non-toggle switching, without any requirement of the sensing procedure and dynamic withdrawal of stress. Therefore, Ref.~\citenum{munir15} misses several key understandings in this respect.
	
 \item  Ref.~\citenum{munir15} states while referring Ref.~\citenum{roy14} that ``We do not use any such construct and retain the energy advantage of SML.'' First, note that Ref.~\citenum{munir15} fails to achieve high switching speed at tolerable error probability for logic design. Ref.~\citenum{munir15} does \emph{not} mention that the proposal in Ref.~\citenum{roy14} achieves high switching speed ($\sim$1ns) at low error probability ($<10^{-4}$) for logic design. The pulse shaping methodology in Ref.~\citenum{munir15} cannot tackle the critical issue behind switching failures as analyzed in Ref.~\citenum{roy14}. 

\end{enumerate}

Apart from making incorrect and misleading statements on Ref.~\citenum{roy14}, Ref.~\citenum{munir15} is erroneous in the following aspects.

\begin{enumerate}

\renewcommand{\theenumi}{\arabic{enumi}}

	\item Ref.~\citenum{munir15} states that ``Since our voltage pulse widths are 1 ns or more, we can neglect effects associated with finite rise and fall times of the stress in response to an abrupt voltage pulse.'' Note that it is shown in literature~\cite{roy11_5,roy13_2,roy11_6} that particularly fall time is important to consider, since it may rotate magnetization out-of-plane in a direction that affects the error probability. Therefore, without a relevant analysis, the comparative basis of using different pulses (e.g., cases 3 and 4 pertain to ramped pulses, while the other cases do not) in Ref.~\citenum{munir15} is untenable. 

	\item Ref.~\citenum{munir15} states that ``Therefore, operating in the dipole dominated region reduces error rate but increases switching delay (because the stress is relatively weak), while operating at stress levels much above the dipole dominated region has the opposite effect.'' The comment that ``operating in the dipole dominated region reduces error rate'' is incorrect, since the error occurs while \emph{withdrawing} the stress. If the magnetization's out-of-plane excursion is in so-called \emph{bad} quadrant, it may be detrimental and magnetization may traverse in the opposite direction, as explained in Ref.~\citenum{roy14}. Therefore finite ramp rate particularly while withdrawing the stress is necessary to consider. Ref.~\citenum{munir15} did not consider finite ramp rate, which is not only unphysical but also very important for relevant quantitative analysis. The dipole coupling makes the potential landscape asymmetric and the degree of asymmetry is same irrespective of stress, which is symmetric. Such statement fails to identify the key issue that dipole coupling makes small asymmetry in the system herein. This is why dipole coupling cannot counter the motion due to out-of-plane excursion, as explained in Ref.~\citenum{roy14}, and thereby Ref.~\citenum{munir15} pertains to a high error probability.
	
\hspace*{5mm}	Also, ``operating at stress levels much above the dipole dominated region has the opposite effect.'' is far from accurate since with high stress level accompanied by finite ramp rate, there can be considerable $\phi$-motion [see Equation~\eref{eq:T_stress}] and precession, increasing the switching delay eventually~\cite{roy11_5,roy13_2,roy11_6}. 
	 
	\item While explaining the peak in Fig.~5, Ref.~\citenum{munir15} says that with the increase of stress ``the influence of dipole interaction is diminished'' etc. This is not a correct explanation since asymmetric dipole coupling is always active alongwith the symmetric stress [see Fig.~\ref{fig:Nanotechnology_26_245202_2015}(i)]. For a correct explanation, one needs to invoke the out-of-plane excursion of magnetization with stress amplitude and stress pulse timing as parameters, as explained in Ref.~\citenum{roy14}.
		
	\item Ref.~\citenum{munir15} states incorrectly that ``Any combination of $a$, $b$ and $t$ that gives a barrier height of 32 kT will yield similar results.'' Note that the demagnetization factors dictate the magnetization dynamics and they depend on the specific values of $a$, $b$ and $t$. For example, switching delay can vary twice or more for the same barrier height~\cite{roy14_5}. This is a basic understanding regarding magnetization dynamics. Also, the barrier height needs to be calculated accurately. Several papers~\cite{biswa13,biswa14,biswa14_2} underestimated the barrier height as much as by 40\% using the assumption $a/b \sim 1$.
	
	\item Ref.~\citenum{munir15} does not provide a correct mathematical procedure to calculate the critical stress in the presence of dipole coupling. It uses a stress of 3 MPa [rather than 2.64 MPa, calculated with a relevant procedure, see Fig.~\ref{fig:Nanotechnology_26_245202_2015}(i)], which does \emph{not} make minimum energy position at the easy axis. 
	
	\item There are less serious but confusing and erroneous issues in Ref.~\citenum{munir15}, e.g., Equations~(7)-(9) should have ``$3\lambda$'' in stead of ``$2\lambda$'' (sometimes $\lambda_s$ is used for $\lambda$), the calculated values of Equations~(8) and (9) correspond to $d=200\,nm$ instead of $d=150\,nm$, in Fig.~4, $\theta$ (instead of $\phi$) is used as azimuthal angle, there are two Fig.~4(c)s in the plot, Figs.~6, 8, and 10 plot $\phi$ from $-\pi$ to $2\pi$, while $\phi$ has a range of only $2\pi$ and it therefore depicts some duplicate and separated distributions incorrectly, also it appears that the last set in the Figs.~6, 8, 10 corresponds to 2 ns rather than 3 ns etc.
	
\end{enumerate}

While there exists a litany of errors in Ref.~\citenum{munir15}, the purpose of this Comment is to primarily point out that Ref.~\citenum{munir15} has made \emph{incorrect} statements on Ref.~\citenum{roy14} as explained earlier: (1) misquoted that Ref.~\citenum{roy14} deals with memory and not logic, (2) did not mention that the proposal in Ref.~\citenum{roy14} can keep the switching speed high $\sim$1 ns alongwith low error probability, unlike the case in Ref.~\citenum{munir15}, and (3) made \emph{misleading} statement that energy dissipation in Ref.~\citenum{roy14} is high. The understandings reflected in Ref.~\citenum{munir15} are incorrect and \emph{not} sound, while predicting the demise of multiferroic nanomagnetic logic in the presence of thermal fluctuations~\cite{munir15,munir14,fasha13,fasha13_2,dsouza14}. 

It should be also pointed out that there is issue with scaling down device dimensions since the analysis performed in Ref.~\citenum{munir15} depends on the \emph{in-plane} asymmetry created by dipole-coupling (between laterally placed nanomagnets), which decreases with scaling. Therefore, in the scaled-down nanomagnets, the error-probability would be even higher than what is mentioned in  Ref.~\citenum{munir15}. There is also similar issue regarding \emph{area consumption} on the proposals in Refs.~\citenum{biswa14,biswa14_2} due to using \emph{lateral} piezoelectric pads, which is tantamount to \emph{micro}-electronics rather than \emph{nano}-electronics~\cite{itrs,nri}, and therefore untenable for meeting practical standard requirement of area density 1 Tb/in$^2$. Both theoretical and experimental works are emerging in this area~\cite{roy_spie_2014} and the issues pointed out here hopefully would play an important role in devising memory and logic in our future information processing systems.

\section*{References}

\providecommand{\newblock}{}


\preprint{AIP/123-QED}

\title{Comment on ``Acoustically assisted spin-transfer-torque switching of nanomagnets: An energy-efficient hybrid writing scheme for non-volatile memory'' [Appl. Phys. Lett. 103, 232401 (2013)]}

\author{Kuntal Roy}
\email{royk@purdue.edu.}
\noaffiliation
\affiliation{Department of Electrical and Computer Engineering, Purdue University, West Lafayette, Indiana 47907, USA}

												
\maketitle

Ref.~\onlinecite{biswa13} claimed an energy-efficient way to write in spin-transfer-torque random access memory (STTRAM) using strain in piezoelectric-magnetostrictive heterostructures. Here, we show that the claim of energy-efficiency is {incorrect} while comparing to both traditional transistors and STTRAMs. Also the energy dissipation calculation due to \emph{global} surface acoustic wave (SAW) is missing a crucial point that will invalidate the claim of energy-efficiency.

\begin{figure}[t]
\centering
\includegraphics{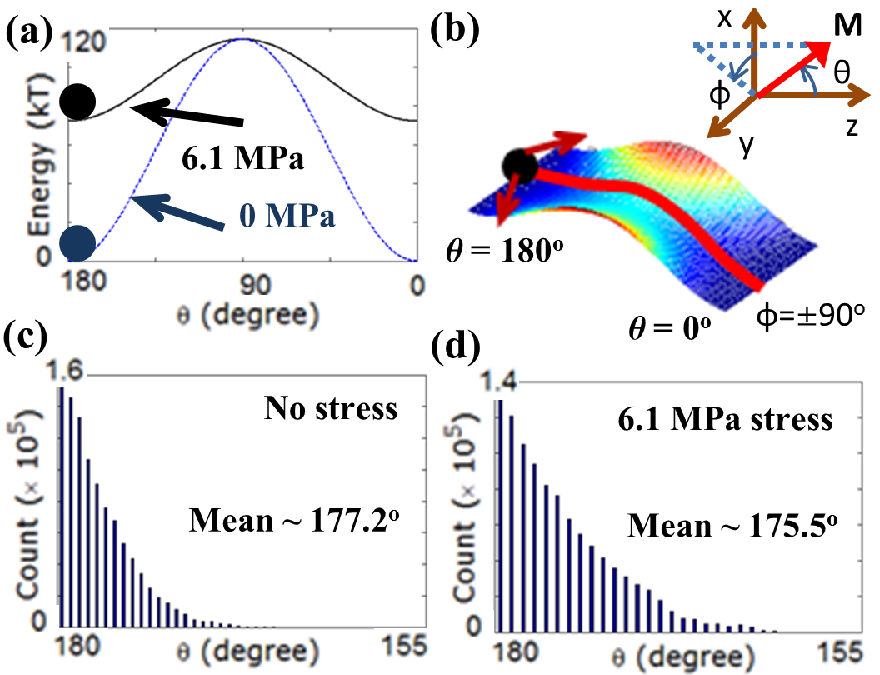}
\caption{\label{fig:APL_103_232401_2013}
(a) Potential landscapes of the nanomagnet with no stress and 6.1 MPa stress. 
(b) Three-dimensional potential landscape showing the deflection of magnetization out of magnet's plane ($y$-$z$ plane, $\phi=\pm90^\circ$).
(c) Room-temperature (300 K) thermal distribution of $\theta$ when no stress is active.
(d) Room-temperature (300 K) thermal distribution of $\theta$ when 6.1 MPa stress is applied.
}
\end{figure}

First, Ref.~\onlinecite{biswa13} follows the idea by Roy presented in the Ref.~\onlinecite{roy10} (was not published due to the issues described here) with the addition of SAW and follows the formulation from Refs.~\onlinecite{roy11_6} and~\onlinecite{roy11_3}. The basic idea was to switch the magnetization by STT along the desired direction when magnetization comes at the hard-plane ($\theta=90^\circ$, Fig.~\ref{fig:APL_103_232401_2013}) upon application of stress.~\cite{roy10} However, thermal fluctuations creates an wide distribution for the time-delay when magnetization reaches at $\theta=90^\circ$.~\cite{roy13_2} This is why STT current needs be kept active for almost half of the duration of switching.~\cite{biswa13} Ref.~\onlinecite{roy13_2} proposed an attractive way to work with only stress by using a sensing element (spin-valve/MTJ) to detect when $\theta$ reaches \emph{around}  90$^{\circ}$ (not at precise juncture as stated by Ref.~\onlinecite{biswa13}).~\cite{roy13_2} 

Figure~3 in Ref.~\onlinecite{biswa13} shows that the energy dissipation with a stress of 6.1 MPa is $5\times10^9$ kT at room-temperature (300 K), which is $\sim$20 pJ. This is 4-5 orders of magnitude higher than that of traditional transistors and therefore {untenable for building nanoelectronics}.~\cite{itrs,nri} 

Also, Fig.~3 in Ref.~\onlinecite{biswa13} shows the energy dissipation when no stress is present (only switching in STTRAM), which is $\sim40\times10^9$ kT = 160 pJ. Note that Ref.~\onlinecite{biswa13} has considered the material Terfenol-D as free layer in STT switching to calculate the  energy dissipation. However, the material that is commonly used for the free layer is CoFeB,~\cite{RefWorks:786,RefWorks:774,RefWorks:33,RefWorks:564} which has Gilbert damping parameter $\alpha$ an order lower~\cite{RefWorks:786,RefWorks:774} than Terfenol-D ($\alpha$ of Terfenol-D is 0.1 used by Ref.~\onlinecite{biswa13}). The critical current of switching is proportional to damping parameter~\cite{RefWorks:786,RefWorks:774} and the energy dissipation is proportional to the square of the switching current. Therefore, Ref.~\onlinecite{biswa13} calculated \emph{incorrectly} the switching current an order higher (23 mA) and energy dissipation about two orders higher ($\sim$160 pJ) for STT switching (Ref.~\onlinecite{RefWorks:786} \emph{correctly} determined switching current $\sim$1 mA and energy dissipation $\sim$1 pJ experimentally). 

Clearly, the comparison with STT switching performed in Ref.~\onlinecite{biswa13} is \emph{incorrect} and actually the energy dissipation in STT switching is 12.5 times lower (in stead of 8 times higher as \emph{incorrectly} claimed by Ref.~\onlinecite{biswa13}) than the hybrid scheme proposed in Ref.~\onlinecite{biswa13}. Note that CoFeB cannot be used as the magnetostrictive nanomagnet instead of Terfenol-D since CoFeB has 30 times lower magnetostriction coefficient than that of Terfenol-D requiring 30 times higher stress (and concomitant higher voltage) to generate the same stress anisotropy subjected to maximum strain allowed in the linear regime of stress-strain relation. 

Also, the energy dissipation calculated due to SAW is missing a crucial point that will make the energy dissipation exceedingly high. Ref.~\onlinecite{biswa13} says that ``SAW is ``global'' and affects every memory cell.'' This is represented in Fig.~\ref{fig:APL_103_232401_2013}(a) that with the application of 6.1 MPa stress the barrier height decreases but it is not sufficient enough to make the potential landscape monostable and cause switching. However, when stress is applied the magnetization \emph{does} rotate. The distribution in Fig.~\ref{fig:APL_103_232401_2013}(d) upon application of stress is wider than the distribution in Fig.~\ref{fig:APL_103_232401_2013}(c) when no stress is active. Therefore, the energy dissipation upon application of stress and removal of stress \emph{must be calculated}, which is \emph{missed} by Ref.~\onlinecite{biswa13}. This energy dissipation turns out to be $\sim$40 kT for one cell. Considering just 1 MB memory, clearly such strategy would dissipate an energy which is quite worse than that of transistors.~\cite{itrs,nri}

Fig.~\ref{fig:APL_103_232401_2013}(b) depicts that magnetization may deflect out of magnet's plane when stress is applied. The torque due to stress acts in the out-of-plane ($\mathbf{\hat{e}_\phi}$ in Fig.~\ref{fig:APL_103_232401_2013}(b)) direction as 
\begin{equation}
\mathbf{T_{E,stress}} = - \mathbf{\hat{e}_r} \times \nabla\,E_{stress} = - (3/2) \, \lambda_s \sigma \Omega sin(2\theta)\, \mathbf{\hat{e}_\phi}, \nonumber
\label{eq:T_stress}
\end{equation}
where $E_{stress}=- (3/2) \, \lambda_s \sigma \Omega cos^2\theta$ is the potential energy due to stress, $(3/2)\lambda_s$ is the magnetostrictive coefficient, $\sigma$ is the stress, $\Omega$ is nanomagnet's volume. Therefore \emph{magnetization rotates out-of-plane} even if the demagnetization factor in the out-of-plane direction is high ($\sim$10 times compared to the in-plane directions based on the chosen dimensions). This out-of-plane excursion has important ramification to increase switching speed,~\cite{roy13_2,roy11,roy11_2,roy11_6} which cannot be conceived if magnetization is \emph{assumed} to reside always on magnet's plane as Ref.~\onlinecite{RefWorks:154} did. This out-of-plane excursion depicted in the Fig.~\ref{fig:APL_103_232401_2013}(b) is considered for the calculation of energy dissipation upon application of stress.

Note that the energy barrier height calculated in Ref.~\onlinecite{biswa13} is underestimated by $\sim$33\% using the \emph{assumption} of $a/b\sim 1$ (while a/b=110 nm/90 nm). In Ref.~\onlinecite{biswa13}, the SAW frequency is 100 MHz and the lateral area of the devices are high (110 $\times$ 90 nm$^2$), which are clearly inferior to traditional transistors.~\cite{itrs,nri} Refs.~\onlinecite{biswa14_2} and~\onlinecite{biswa14} are even more higher area consuming due to using lateral piezoelectric pads and also switching delay is high by about an order due to ``toggle'' switching.~\cite{biswa14_2}

%